\documentclass[conference,compsoc]{IEEEtran}
\usepackage{graphicx}
\usepackage{soul}
\usepackage{color}
\usepackage{listings}
\usepackage{float}
\usepackage{wrapfig} 
\usepackage{tabularx}
\usepackage{algorithm}
\usepackage{algpseudocode}
\usepackage{pifont}
\usepackage{url}
\usepackage{enumerate}
\usepackage{enumitem}  
\usepackage{amsmath,amsfonts,amssymb}
\usepackage{amsthm}
\usepackage[utf8]{inputenc}
\usepackage[english]{babel}

\usepackage{subfigure}		
\usepackage{soul}					
\usepackage{qtree}
\usepackage[bottom]{footmisc}   

\theoremstyle{definition}
\newtheorem{definition}{Definition}[section]
\newtheorem{example}[definition]{Example}
\theoremstyle{plain}

\theoremstyle{remark}

\ifCLASSOPTIONcompsoc
  \usepackage[nocompress]{cite}
\else
  \usepackage{cite}
\fi
\ifCLASSINFOpdf
\else
\fi
\begin{document}
%
\title{Applying High-Performance Bioinformatics Tools for Outlier Detection in Log Data}


\author{\IEEEauthorblockN{Markus Wurzenberger, Florian Skopik,
		Roman Fiedler}
\IEEEauthorblockA{Austrian Institute of Technology, Center for Digital Safety and Security\\
	Donau-City-Strasse 1,
	1220 Vienna, Austria\\
	firstname.lastname@ait.ac.at}
\and
\IEEEauthorblockN{Wolfgang Kastner}
\IEEEauthorblockA{Vienna University of Technology\\
	Treitlstrasse 3,
	1040 Vienna, Austria\\
	k@auto.tuwien.ac.at}
}

%


\maketitle

\begin{abstract}
Most of today's security solutions, such as security information and event management (SIEM) and signature based IDS, require the operator to evaluate potential attack vectors and update detection signatures and rules in a timely manner. However, today's sophisticated and tailored advanced persistent threats (APT), malware, ransomware and rootkits, can be so complex and diverse, and often use zero day exploits, that a pure signature-based blacklisting approach would not be sufficient to detect them. Therefore, we could observe a major paradigm shift towards anomaly-based detection mechanisms, which try to establish a system behavior baseline -- either based on netflow data or system logging data -- and report any deviations from this baseline. While these approaches look promising, they usually suffer from scalability issues. As the amount of log data generated during IT operations is exponentially growing, high-performance analysis methods are required that can handle this huge amount of data in real-time. In this paper, we demonstrate how high-performance bioinformatics tools can be applied to tackle this issue. We investigate their application to log data for outlier detection to timely reveal anomalous system behavior that points to cyber attacks. Finally, we assess the detection capability and run-time performance of the proposed approach.
\end{abstract}

\section{Introduction}
\label{sec:introduction}
Many of today's ICT security solutions promise an automatic detection (and even mitigation) of malicious behavior. They apply complex detection schemes and heuristics -- and massive data exchange across systems, organizations and even countries: Threat Intelligence is the new hype. But still, effective monitoring of a technical infrastructure is the most essential phase of security incident handling within organizations today.

To establish situational awareness, it is indispensable for organizations to have a thorough understanding about what is going on in their network infrastructures. Therefore, clustering techniques are very effective tools for periodically reviewing rare events (outliers) and checking frequent events by comparing cluster sizes over time (e.g., trends in the number of requests to certain resources). Furthermore, a methodology and supporting tools to review log data and to find anomalous events in log data are needed. Existing tools are basically suitable to cover all these requirements, but they still suffer from some essential shortcomings. Most of them, such as SLCT \cite{SLCT2003}, implement word-based matching of log entries, but for example do not identify synonyms which only differ in one character, such as `php-admin' and `phpadmin'; or consider similar URLs as completely different words, although they have the same meaning. Hence, the implementation of character-based matching with comparable speed such as word-based matching is necessary. Furthermore, existing tools are often notcapable of processing large log files to extract cluster candidates over months and to perform gradual logging, which can be applied to generate a base corpora to identify how and where log clusters are changing over the time.

In the domain of bioinformatics, various methods have been developed to analyze and study the similarity of biologic sequences (DNA, RNA, amino acids), group similar sequences and extract common properties \cite{wang_data_2006}. The algorithms that implement these features need to fulfill some fierce requirements -- similarly important to process log data:
\begin{itemize}
	\item \textit{Adequate digital representation:} Biologic sequences must be represented as data streams in an appropriate format, i.e., no information must be lost, but the format should be as simple as possible.
	\item \textit{Dealing with natural variations:} The dependency between a segment of a sequence and  a certain biologic function (implemented by this segment) is sometimes not strict (or obvious). This means natural variations need to be accounted for and a certain degree of fuzziness in the input data accepted.
	\item \textit{Dealing with artificial inaccuracies:} The process of recording long and complex biologic sequences causes inevitable inaccuracies and small errors. The negative influence of those (artificially introduced) variations in the following analysis phase should, however, be kept to an absolute minimum.
	\item \textit{Dealing with massive data volumes:} Since biologic data sequences are (even to represent simple functions) very complex, algorithms need to deal with these large amounts of data usually by (i) being scheduled in parallel and (ii) accepting certain inaccuracies caused by this non-sequential processing.
\end{itemize}	

In general, all these requirements also apply to modern log data processing as (i) data needs to be processed extremely fast (this means depending on the application approximately in real time); (ii) data analysis needs to be scheduled in parallel in order to scale; and (iii) the process needs to accept certain inaccuracies and errors that occur due to conversion errors from varying character encodings, and slight differences in configurations and output across software versions. Furthermore, these tools aim at processing character sequences without taking into account their semantic meanings.

As a consequence, if mentioned tools are not applied to biologic sequences but to re-coded (converted) digital sequences, such as log data (or even malware code), all of the unique properties of these algorithms can be exploited directly, without the need to design and implement complex tools again.

In this paper, we define a method for re-coding log data into the alphabet used for representing canonical amino acid sequences. This allows us to apply high-performance bioinformatics tools to cluster log data. Based on the clustering we perform outlier detection analysis to discover anomalous and erratic behavior. Furthermore, we investigate the applicability and feasibility of our approach in a real setting by simulating a scenario of an attack and evaluate the proposed approach. Finally, we provide an outlook for further applications of the novel model beyond straightforward outlier detection, such as time series analysis to discover anomalous trends.


The remainder of the paper is structured as follows. Sect. \ref{sec:relatedwork} outlines important background and related work. Then, Sect. \ref{model} describes the overall model for discovering outliers in log data. Section \ref{sect:recoding} elaborates on the re-coding model, which transforms log line content into a representation which can be understood by bioinformatics tools. After that, we describe how log lines are compared and clustered in Sect. \ref{clustering} and the detection of outliers in Sect. \ref{sec:outlier}. Section \ref{sec:evaluation} demonstrates the application of our approach and evaluates the feasibility in a realistic setting. Finally, Sect. \ref{sec:conclusion} concludes the paper.
\section{Background and Related Work}
\label{sec:relatedwork}

In the domain of cyber security, logging and log data management are of high importance and improve visibility and security intelligence for computer networks. Thus, log data is a source for security- and computer network analysis tools such as anomaly detection \cite{chandola2009anomaly} and intrusion detection systems \cite{axelsson2000intrusion} that identify anomalous system behavior. In this paper, we focus on the application of the proposed model in the domain of cyber security and concentrate on outlier detection for anomaly and intrusion detection. Various outlier detection methods are discussed in \cite{hodge2004survey}.

The main technique we apply in our model for detecting outliers and for creating a computer network's situation picture is clustering. Many different clustering approaches and algorithms are surveyed in \cite{berkhin2006survey}. For clustering log lines density and distance based approaches can be applied. Simple Logfile Clustering Tool (SLCT) \cite{SLCT2003} is an example for a density based clustering algorithm especially developed for clustering log data. But the proposed model focuses on distance based algorithms.

One of the first metrics to compare two sequences of any kind of symbols, hence also log lines, was the Hamming distance \cite{hamming1950error}, which bases on the number of mismatches and therefore can only be applied to sequences of the same length. A further development of thi metric is the Levenshtein or edit distance \cite{levenshtein1966binary}, which also recognizes insertions and deletions and therefore enables comparison of sequences of different size.

Most bioinformatics tools for clustering amino acid or DNA sequences apply a modified version of the previously mentioned Levenshtein distance. In the case of amino aecid sequences, the number of occurring unique symbols reduces from 256 (in UTF-8 code) to 20 (canonical amino acids). Furthermore, these algorithms make use of the knowledge that there exist empirical statistics that one amino acid naturally can evolve to another one over time. The most popular scoring matrices representing these relations are the PAM (point accepted mutation) matrix and the BLOSUM (blocks substitution matrix) matrix \cite{henikoff1992amino}.

There exist a couple of sequence alignment algorithms that compare two amino acid sequences. Therefore, a distinction is made between global alignment, where all symbols of two sequences are compared, such as the Needleman-Wunsch algorithm \cite{needleman1970general}, the Hirschberg algorithm \cite{hirschberg1975linear} and the Gotoh algorithm \cite{gotoh1982improved} and local alignment, where just a subsequence is compared, such as the Smith-Waterman algorithm \cite{smith1981identification}. There exist also fast heuristic algorithms such as FASTA \cite{pearson1988improved} and BLAST \cite{altschul1990basic} to produce alignments.

Various algorithms for clustering amino acids exploit sequence alignment. Some examples are CD-HIT \cite{li2002tolerating}, CLUSTAL \cite{higgins1988clustal} and UCLUST \cite{edgar2010search}. CD-HIT also applies a powerful short word filter, which significantly improves the performance of the clustering algorithm.

%
%
%
%
%
%

\section{Model for Applying Bioinformatics Clustering Tools on Log Data}\label{model}
The following section defines the theoretical model for applying high-performance bioinformatics tools for clustering computer log data, which we already motivated in \cite{wurzenberger2016discovering}. The proposed modular model comprises several steps from re-coding log data to the alphabet used for describing amino acid sequences to interpretation and analysis of the output for cyber security application:
\begin{enumerate}[label=(\roman*)]
	\item collect log data,
	\item homogenize log data,
	\item re-code and format log data,
	\item compare pairs of log lines according to their similarity,
	\item cluster log lines,
	\item retranslate data,
	\item detect outliers and analyse time series.
\end{enumerate}

\begin{figure}[tb]
	\centering
	\includegraphics[width=\linewidth]{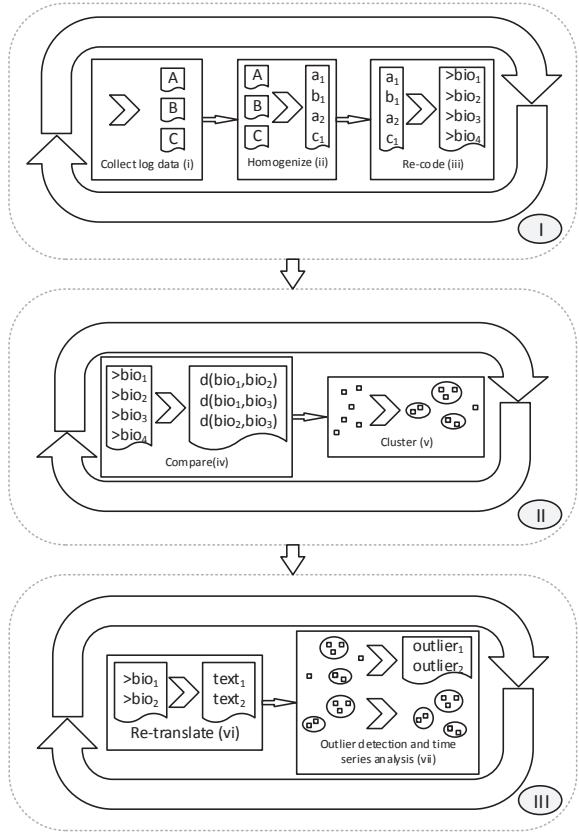}
	\caption[Process Description]{Visualization of the model for applying high-performance bioinformatics tools for the application in the domain of cyber security.}
	\label{fig:process}
\end{figure}

Figure \ref{fig:process} visualizes the proposed model. As the figure shows, the model can be roughly divided into three blocks which are sequentially repeated. Block I covers the process of re-coding log data into a format, which can be exploited by bioinformatics tools. First, in step (i) log data from different sources of the monitored network is collected. When analyzing log data from different sources usually the data shows some differences in the format. For example, main properties such as time stamps are represented in different formats. Therefore, step (ii) is required to homogenize the data. A common time stamp format is important to order log lines chronologically when combining data from different sources. In step (iii), the homogenized data is re-coded from UTF-8 (256 symbols)  to the alphabet describing the canonical amino acids (20 symbols).

In block II, bioinformatics tools are applied to the re-coded data. During step (iv) the re-coded log lines are compared and a distance $d$ between all pairs of lines is calculated. Therefore, a sequence alignment algorithm for amino acid sequences is applied to the data. Based on the calculated distance in step (v), the log lines then are clustered.

Block III implements the security analysis component of the proposed model. First in step (vi), a reverse look up function is used to re-translate the log lines from the alphabet describing canonical amino acids into UTF-8 encoded log data, which is readable for human users. Finally in step (vii), an outlier detection and time series/trend analysis is performed to detect on the one hand rare events and on the other hand changes in the common system behavior. Both can be caused by cyber attacks or invaders, as well as misconfiguration and erratic system behavior. In the remaining paper we focus on outlier detection.
\section{Re-coding Model}\label{sect:recoding}
Log data from ICT systems is usually modeled in human-readable textual form. Therefore, before tools from the domain of bioinformatics can be applied to it, step (iii) has to be carried out, i.e., re-coding the log data using the alphabet used for representing amino acids and converting it into a format, which can be exploited by the applied tools.

A basic unit of logging information, e.g., one line for line-based logging, or one XML-element, is called a textual log atom $L_{text}$ which consists of a series of symbols $s$ -- typically letters and numbers (Eq. \ref{eq:log-atom}).The used alphabet to represent log data consists (in most cases) of UTF-8 encoded characters (256 different symbols) of 8 bit size. In the following $A_{UTF-8}$ refers to this alphabet.

\begin{equation}
\label{eq:log-atom}
L_{text} = \left\langle s_1 s_2 s_3 \dots s_n \right\rangle \text{ where } s_i \in A_{UTF-8}
\end{equation}

But data represented in this format is unsuitable as input to bioinformatics tools. Those tools require input (biologic sequences) encoded with symbols of the alphabet $A_{bio}$ (Eq. \ref{eq:bio-alphabet}) defined for amino acid or DNA sequences. This alphabet consists of 20 symbols only, which represent the 20 canonical amino acids.

\begin{equation}
\label{eq:bio-alphabet}
A_{bio} = \text{\{A,C,D,E,F,G,H,I,K,L,M,N,P,Q,R,S,T,V,W,Y\}}
\end{equation}

A re-coding function takes an input stream encoded as UTF-8 data and transforms it into a representation $L_{bio}$ (Eq. \ref{eq:log-bio}) that is processable by bioinformatics tools.

\begin{equation}
\label{eq:log-bio}
L_{bio} = \left\langle s_1 s_2 s_3 \dots s_m \right\rangle \text{ where } s_j \in A_{bio} 
\end{equation}

In the simplest case, this transformation is a straight forward bijective mapping, where one $A_{UTF-8}$ symbol is represented by two symbols from $A_{bio}$. However, for data where certain larger blocks frequently appear, those whole blocks (e.g., server names or IP addresses) could be replaced with a single symbol. This would effectively allow compression of data. Even further information loss could be -- depending on the application use case -- acceptable. For instance, frequently appearing symbol blocks could be replaced through applying a more intelligent, but just one way mapping, e.g., not a whole IP address but just the last byte or the address' cross sum could be translated to $A_{bio}$. Another example are paths (from Web server logs), where each component of a path could be translated through hashing into single symbols of $A_{bio}$. Furthermore, symbols can be grouped by type, so that for example all separators such as '/', ';' or spaces can be replaced by one specific element of $A_{bio}$.

Even more complex re-coding schemes are possible, e.g., after identifying dynamic and static parts of log lines with simple log line clustering tools (such as SLCT \cite{SLCT2003}), more symbols could be spent on the variable parts of log lines (those with higher information entropy) and less symbols (or no symbols at all) on the rather static parts.

One simple - but effective - method for re-coding log data into $A_{bio}$ is described in details in the following. In order to re-code $L_{text}$ into $L_{bio}$, a simple and straight forward solution is to convert each $s_i \in L_{text}$ into two\footnote{Since the size of the alphabet of $L_{text}$ is larger (256 elements) than that of $L_{bio}$ (20 elements), one $s_i \in L_{text}$ has a much higher entropy than one $s_j \in L_{bio}$.} corresponding $s_j \in L_{bio}$ symbol by symbol (without any loss of information). For this purpose, each symbol in $L_{text}$ (i.e., the single letters of the words in a log line) is converted to its numerical representation in UTF-8. The result of this operation is $L_{utf}$ (Eq. \ref{eq:log-utf8}).

\begin{equation}
\label{eq:log-utf8}
L_{utf} = \left\langle a_1,a_2,a_3 \dots a_n \right\rangle \text{ where } a_i \in \{0,\ldots,255\} 
\end{equation}

In a second step, each numerical value $a_i\in L_{utf}$ is converted into two symbols of the alphabet $A_{bio}$. Since the size of this alphabet is always $20$, a straight forward solution (and in order to use the whole possible input range) is to divide each $a_i\in L_{utf}$ by $20$, and additionally keep the rest of this division. Eventually, both results $s_1$ (the result of the integer division) and $s_2$ (the rest of the division) are mapped via a simple conversion table (see Tab. \ref{tab:biosymbols}) to $A_{bio}$. Concatenating all these symbols in a single stream effectively produces $L_{bio}$ -- the input to alignment and clustering tools from the domain of bioinformatics.

\begin{table}[!ht]
	\centering
	\scriptsize
	\caption{\label{tab:biosymbols} Bio Alphabet Symbol Mapping.}	
	\begin{tabular}{|l|c|c|c|c|c|c|c|c|c|c|}  \hline
		Number: &0  &1  &2  &3  &4  &5  &6  &7  &8  &9 \\ \hline
		Symbol: &A  &C  &D  &E  &F  &G  &H  &I  &K  &L \\ \hline \hline
		Number: &10 &11 &12 &13 &14 &15 &16 &17 &18 &19 \\ \hline
		Symbol: &M  &N  &P  &Q  &R  &S  &T  &V  &W  &Y \\ \hline
	\end{tabular}			
\end{table}

The re-coding process is further described in Alg. \ref{alg1}. There, the symbol $\oplus$ extends the collection on the left side with the symbol on the right side. The function \texttt{utf2num} looks up the decimal symbol number in a standard UTF-8 table (e.g., the letter `A' corresponds to the number 65). The function \texttt{num2bio} looks up the letter representation of the numbers 0 to 19 (according to Tab. \ref{tab:biosymbols}).

\begin{algorithm}
	\scriptsize
	\caption{\label{alg1}Re-Coding $L_{text}$ into $L_{bio}$ }
	\begin{algorithmic}[1]
		\State $L_{bio}\gets \emptyset $
		\State $L_{utf}\gets \emptyset $
		\ForAll{$s_i\in L_{text}$} 
		\State $L_{utf} \gets L_{utf} \oplus \text{utf2num}(s_i)$ \EndFor
		
		\ForAll{$a_i\in L_{utf}$} 
		\State $s_1 \gets a_i/20$
		\State $s_2 \gets a_i\%20$
		\State $L_{bio} \gets L_{bio} \oplus \text{num2bio}(s_1) \oplus \text{num2bio}(s_2)$ 
		\EndFor
	\end{algorithmic}
\end{algorithm}

 A simple option to reduce/compress the mount of data needed to represent one log line by $50\%$ is, instead of representing each $s_i \in L_{text}$ by two $s_j \in L_{bio}$ (cf. $L_{bio}^{full}$ in Tab \ref{tab:rec-ex}), to omit the leading character $s_1$ (cf. Alg. \ref{alg1}). This one has less entropy compared to the trailing $s_2$ (cf. Alg. \ref{alg1}), because if all $256$ letters of $A_{UTF-8}$ are occurring in the considered data only the first $12$ letters of $A_{bio}$ are used to represent $s_1$. Since usually less than $100$ symbols of $A_{UTF-8}$ occur in a realistic dataset and these symbols are in general numerically represented in the same region, for example $a_i\in\{51,\ldots,150\}$, which reduces the possible options for $s_1$ to at most $5$. Hence, only a quarter of all possible options is used to describe $s_1$. As a consequence, $s_1$ stores less information than $s_2$. Furthermore, while one $s_1$ occurs in the description of $20$ symbols, one $s_2$ maximally represents $5$ symbols of $A_{UTF-8}$. Finally, when omitting $s_1$ still the combination of $s_j\in A_{bio}$ in $L_{bio}$ raises the entropy of every $s_2$ obtained from Alg. \ref{alg1}. Hence, the result is that the length of $L_{bio}$ can effectively be cut to a half by accepting a "small" ambiguity (cf. $L_{bio}$ in Tab \ref{tab:rec-ex}). In the remaining paper, we always apply this method for recording $L_{text}$ into $L_{bio}$.

\begin{table}[!ht]
	\centering
	\scriptsize
	\caption{\label{tab:rec-ex} Step-by-Step recoding example.}	
	\begin{tabular}{|l|c|c|c|c|c|c|c|c|c|c|}  \hline
		$L_{text}$:&1  &9  &2  &.  &1  &6  &8  &.  &... \\ \hline
		$L_{utf}$: &49 &57 &50 &46 &49 &54 &56 &46 &... \\ \hline
		$L_{bio}^{full}$: &DL  &DV  &DM  &DH  &DL  &DR  &DT  &DH  &... \\ \hline
		$L_{bio}$: &L  &V  &M  &H  &L  &R  &T  &H  &... \\ \hline
	\end{tabular}			
\end{table}

To complete step (iii), the data has to be transformed into the correct format. Most of the bioinformatic tools require data in the FASTA format, which has been introduced by Lipman and Pearson \cite{pearson1988improved}. An example for this format is given in Listing \ref{lst:fastaformat}. As described later, the header required by the FASTA format can be used to store information for the re-translation implemented by step (vi).

\begin{figure}[htb]
	\begin{lstlisting}[mathescape,language=XML,columns=fullflexible,keepspaces=true,frame=tb,basicstyle=\tiny\ttfamily,numbersep=3pt,caption=Example for two sequences is FASTA format.,captionpos=b,stringstyle=\ttfamily,label={lst:fastaformat},breaklines=true,keywordstyle=\bfseries,morekeywords={}]
> 0x
LVMHLRTHLVLHPPGPGPNNKIECPIMKLPWKKWMMWKQPEKKKKQPRNLFPI...
> 1x
LVMHLRTHLVLHPPGPGPNNKIECPIMKLPWKKWMMWMQPEKKKKQPRNLFPI...
	\end{lstlisting}
\end{figure}

\section{Comparing and Clustering of Log Data with Bioinformatics Tools}\label{clustering}
A promising extension of bioinforamtics tools, such as CD-HIT \cite{li2002tolerating}, CLUSTAL \cite{higgins1988clustal} and UCLUST \cite{edgar2010search} for log analysis in order to increase the accuracy of results (identifying regions of similarity etc.) is the application of bio-clustering on re-coded log data. Since bio-clustering applies methods to group similar sequences which are fundamentally different from the commonly applied text mining approaches, using bio-clustering can considerably improve the quality of results. The reason for this improvement is manifold. First, many common correlation algorithms require decent knowledge about the syntax and semantics of the input data. But, however this is not realistic for logging data from different systems. Second, many text mining and clustering algorithms lack the required degree of uncertainty when processing log data. For instance, if two words differ by just one letter, they are usually considered as completely different in the clustering process, because for text mining, synonym tables are more appropriate. This is however not true for log data, where text junks, such as `php-admin' and `phpadmin' should be considered similar, if not almost equal. Alignment algorithms from the domain of bioinformatics assist by using a different metric to measure word similarity which eventually improves the effectiveness. Third, most text mining algorithms do not handle special characters adequately, as they have different meanings in regular text and log messages. For instance `../../etc/passwd' is hard to process, as `.' and `/' are considered natural delimiters in written texts, but not in logs. Additionally, certain log sequences, e.g., paths `../../etc/passwd' and `../../../etc/passwd', have considerably different meaning, however look similar for text mining algorithms. Through applying deletions and insertions in the bio-representations, these properties are adequately handled by bioinformatics tools, and thus, distances calculated accordingly. In the following, we describe how alignment algorithms from the domain of bioinformatics work for comparing two into $A_{bio}$ re-coded log lines and how bio-clustering tools can be used for grouping log data.

\subsection{Pairwise Log Line Comparison}\label{sect:compare2lines}
Sequence alignment algorithms, which are applied in step (iv), cf. Sect. \ref{model} to compare two log lines form the base for most bio-clustering tools. Some examples for such algorithms are given in Sect. \ref{sec:relatedwork}. Alignment algorithms use a scoring function $d$ to calculate the distance between two sequences. When comparing two sequences $L_{bio}^A$ and $L_{bio}^B$ element by element, there can occur three possible cases:
\begin{enumerate}
	\item \textbf{mismatch}: symbol $s_j^A$ was replaced by symbol $s_j^B$,
	\item \textbf{deletion}: symbol $s_j^A$ was removed in $L_{bio}^B$,
	\item \textbf{insertion}: symbol $s_j^B$ was inserted in $L_{bio}^B$.
\end{enumerate} 

The alignment between two amino acid sequences is always built under the assumption that $L_{bio}^A$ and $L_{bio}^B$ have common ancestors, i.e., they are homologous. This means in the end the alignment which refers to the highest similarity is chosen \cite{mount_bioinformatics:_2004}. How similar two amino acid sequences are is specified by a similarity score. The predefined score for a match is usually constant. In most cases, the score for a mismatch depends on the probability that $s_j^A$ can evolve to $s_j^B$ over time. These probabilities are based on empirical statistics and represented in a $20\times 20$ lower triangular matrix, which is called scoring matrix. The score for a gap caused by deletions or insertions is also predefined and can depend on the size of the gap, or if a gap is opened or just extended. The simplest definition for a scoring function $d$ relies on unit costs. In the following, we apply this simple scoring system, which does not take into account that $s_j^A$ could evolve to $s_j^B$ by a specific probability:
\begin{align}\label{costfct}
\begin{aligned}
&d(s_j^A,s_j^B)=\begin{cases}
1  & \text{if }s_j^A=s_j^B,\\
-1, & \text{is }s_j^A\ne s_j^B,
\end{cases}\\
&d(s_j^A,-)=-1 \quad\text{deletion},\\
&d(-,s_j^B)=-1 \quad\text{insertion}.
\end{aligned}
\end{align}

When comparing two amino acid sequences, there are usually various options to build the alignment. In our model, since the sequences are considered as homologous, the alignment with the highest score is chosen, because a higher score suggests a higher similarity. Example \ref{examplescores} shows how the optimal alignment is chosen.
\begin{example}\label{examplescores}
	Given two amino acid sequences $L_{bio}^A=\texttt{GAC}$ and $L_{bio}^B=\texttt{GC}$. We assume that $L_{bio}^A$ and $L_{bio}^B$ are homologous. As scoring function serves $d$ defined in Eq. (\ref{costfct}). Table \ref{tab:scoresexample} summarizes the possible alignments. Here option (iii) would be the optimal alignment since it has the highest score.
	\begin{table}
		\center
		\scriptsize
		\begin{tabular}[c]{| c | c | c |}
			\hline
			\textbf{option} & \textbf{alignment} & \textbf{score} \\ \hline\hline
			(i) & \begin{tabular}[x]{@{}c@{}}\texttt{GAC}\\\texttt{GC-}\end{tabular}  & $1-1-1=-1$ \\  \hline
			(ii) & \begin{tabular}[x]{@{}c@{}}\texttt{GAC--}\\\texttt{---GC}\end{tabular} & $-1-1-1-1-1=-5$ \\ \hline 
			(iii) & \begin{tabular}[x]{@{}c@{}}\texttt{GAC}\\\texttt{G-C}\end{tabular} & $1-1+1=1$\\ \hline
		\end{tabular}
		\caption{\label{tab:scoresexample}Scores for Example \ref{examplescores}}
	\end{table}
\end{example}

In the proposed model, the similarity, between the two amino acid sequences can be calculated as the ratio between the number of identical symbols in the alignment and the length of the alignment as shown in Eq. (\ref{similarity}). Equation (\ref{similarity}) is a normalized version of the inverted Lvenshtein distance \cite{levenshtein1966binary}, i.e., the identical symbols are calculated instead of the number of changes. In the case of Example \ref{examplescores}, the similarity for option (iii) would be approximately $66,66\%$.
\begin{align}\label{similarity}
similarity = \frac{identicalSymbolsAlign(L_{bio}^A,L_{bio}^B)}{lengthOfAlign(L_{bio}^A,L_{bio}^B)}
\end{align}

Listing \ref{lst:full-ex} shows a full example of the comparison of the two bio-encoded sequences $L_{bio}^A$ and $L_{bio}^B$ from Listing \ref{lst:fastaformat}, generated with the BLAST tool \cite{altschul1990basic}. The output of this tool is the alignment of $L_{bio}^A$ and $L_{bio}^B$ (see \texttt{Query} and \texttt{Subject} depicted by Listing \ref{lst:full-ex}). The result is \texttt{Algn}, where gaps are inserted between the residues so that identical or similar characters are aligned in successive columns. In case there is a bijective mapping back to the original data $L_{text}$, the original $L_{text}^A$ and $L_{text}^B$ can be depicted aligned using an inverse function (refer to Listing \ref{lst:full-ex}). Eventually, the differences between the original input lines are marked with either `X', which means different symbols on the respective positions in $L_{text}^A$ and $L_{text}^B$; or `-' which means that there is a gap and input stream \texttt{Query} could not be aligned to \texttt{Subject} for the symbols on this position.

\begin{figure*}[t]

	\begin{lstlisting}[mathescape,language=XML,columns=fullflexible,keepspaces=true,frame=tb,basicstyle=\tiny\ttfamily,numbersep=3pt,caption=Full example from real data: The first block shows the input in textual form; the second block the bio-encoded sequences; the third block the aligned output in bio-representation; the fourth block the aligned version in text representation and the fifth block outlines the differences (`X' means different symbols in the input streams and `-' means gaps).,captionpos=b,stringstyle=\ttfamily,label={lst:full-ex},breaklines=true,keywordstyle=\bfseries,morekeywords={}]
	$L_{text}^A$: 192.168.191.4 - - [30/Sep/2014:00:22:05 +0000] "GET /login_page.php HTTP/1.1" 200 3307 "-" "Zabbix monitoring"               
	$L_{text}^B$: 192.168.191.4 - - [30/Sep/2014:00:22:25 +0000] "GET / HTTP/1.1" 200 5300 "-" "Zabbix monitoring"
	
	$L_{bio}^A$:  LVMHLRTHLVLHPPGPGPNNKIECPIMKLPWKKWMMWKQPEKKKKQPRNLFPIKNEGMSPVECHPFPPPFFAILHLRPMKKPNNKSPRGRPRMVWWGAPLNMGTNRGMER	
	$L_{bio}^B$:  LVMHLRTHLVLHPPGPGPNNKIECPIMKLPWKKWMMWMQPEKKKKQPRNLFPIPPFFAILHLRPMKKPQNKKPRGRPRMVWWGAPLNMGTNRGMER
	
	Query: LVMHLRTHLVLHPPGPGPNNKIECPIMKLPWKKWMMWKQPEKKKKQPRNLFPIKNEGMSPVECHPFPPPFFAILHLRPMKKPNNKSPRGRPRMVWWGAPLNMGTNRGMER
	Algn:  LVMHLRTHLVLHPPGPGPNNKIECPIMKLPWKKWMMW QPEKKKKQPRNLFPI              PPFFAILHLRPMKKP NK PRGRPRMVWWGAPLNMGTNRGMER
	Sbjct: LVMHLRTHLVLHPPGPGPNNKIECPIMKLPWKKWMMWMQPEKKKKQPRNLFPI--------------PPFFAILHLRPMKKPQNKKPRGRPRMVWWGAPLNMGTNRGMER
	
	Query: 192.168.191.4 - - [30/Sep/2014:00:22:05 +0000] "GET /login_page.php HTTP/1.1" 200 3307 "-" "Zabbix monitoring"
	Algn:  192.168.191.4 - - [30/Sep/2014:00:22:X5 +0000] "GET /               HTTP/1.1" 200 X30X "-" "Zabbix monitoring"
	Sbjct: 192.168.191.4 - - [30/Sep/2014:00:22:25 +0000] "GET /-------------- HTTP/1.1" 200 5300 "-" "Zabbix monitoring"
	
	Diff:                                       X               --------------               X  X
	\end{lstlisting}
\end{figure*}
\subsection{Log Line Clustering}
Step (v), cf. Sect. \ref{model}, clustering log data is based on the previously defined alignment of two bio-encoded log lines. By re-coding a whole log data set and subsequent pairwise comparison of bio-encoded log lines through sequence alignment as shown before, distances can be determined by calculating the similarity of two sequences (cf. Eq. (\ref{similarity})). Clustering tools then try to cluster the bio-encoded sequences in a way so that the distances between any two cluster members $c_i \in C$, $c_j \in C$ is lower than the distance to the next cluster center. This analysis can be performed with various existing bio-clustering tools, such as the prominent CD-HIT \cite{li2002tolerating}. CD-HIT first applies an efficient and fast short word filter. If a sequence is considered as similar to the representative sequence of a cluster, the alignment and the exact similarity is calculated. Based on this, the algorithm decides if the sequence corresponds to the cluster or not.

For further analysis of the clustering output, the sequences have to be re-translated into understandable text -- step (vi). Therefore, the FASTA format provides the possibility to store the position of a log line in the original log file in the header (cf. Listing \ref{lst:fastaformat} \texttt{> 0x} and \texttt{> 1x}). Using this information, it is possible to look up the corresponding log line for each bio-encoded sequence in the input log file. 
\section{Outlier Detection}\label{sec:outlier}
The following section briefly deals with step (vii) -- outlier detection for detecting anomalies. Outlier detection aims at identifying so-called point anomalies \cite{chandola2009anomaly}. These outliers are clusters with just a few elements and/or usually a large distance to other clusters, which define the normal state of an network environment. In case of log data, outlier clusters include rare or atypically structured events (log entries). Those outliers are log entries, that require further investigations. Eventually, the previously defined model allows to apply high-performance bioinformatics tools on log data to cluster log lines. During the re-translation from $A_{bio}$ to $A_{UTF-8}$ the clusters can be sorted by size to detect clusters of small size, which represent the outliers. Since it is also possible to generate a representative alignment for every cluster, i.e., to generate a multiple sequence alignment accounting for all log lines assigned to one cluster, the function defined in Eq. (\ref{similarity}) can be used to calculate the distance between all obtained clusters. Hence, it is possible to discover the clusters, with the largest distance to the group of clusters describing the typical system behavior of a network environment.
\section{Evaluation}
\label{sec:evaluation}

The following section deals with the evaluation of the proposed approach for outlier detection in computer networks. The evaluation shows on the one hand the detection capability of our model and on the other hand evaluates the run-time performance of the approach. The section is structured as follows: First, we describe the set-up of the evaluation environment and the configuration of the different components of the model. Then, we introduce the use case on which the evaluation of the detection capability bases and the test data we used for the evaluation. Finally, the evaluation results are discussed.

\subsection{Evaluation environment set-up and model configuration}\label{evalconf}
As test environment, we used a workstation with an Intel Xeon CPU E5-1620 v2 at 3.70GHz 8 cores and 16 GB memory, running Ubuntu 16.04 LTS operating system.

The implementation of the model consists of three main parts. First, we use a python script to re-code the log data from UTF-8 code to the alphabet of canonical amino acids. Therefore, we apply the method described in Alg. \ref{alg1}. During the evaluation, we compare two different methods for re-coding log data. Once we translate the log lines to $L_{bio}^{full}$ (translation without loss of information) and once we comprise the data by re-coding to $L_{bio}$ (translation, which compresses the amount of data by just storing the second character with higher entropy, and therefore leads to a loss of information), as shown in Tab. \ref{tab:rec-ex}.

Second, CD-HIT \cite{li2002tolerating} is used for clustering the re-coded log data. Since we have not evaluated an optimal configuration for the scoring matrix so far and the predefined matrix applied by CD-HIT is using properties of amino acid sequences, we modified the downloaded C++ scripts\footnote{http://weizhongli-lab.org/cd-hit} and defined the scoring function as shown in Eq. (\ref{scoringeval}). In this formulation, the gap symbolizes an insertion or deletion.
\begin{align}\label{scoringeval}
\begin{aligned}
&d(s_j^A,s_j^B)=\begin{cases}
6  & \text{if }s_j^A=s_j^B,\\
-5, & \text{is }s_j^A\ne s_j^B
\end{cases}\\
&d_{\text{open gap}}=-11\\
&d_{\text{extend gap}}=-1 
\end{aligned}
\end{align}
Furthermore, we configured the algorithm, so that every log line is added to the cluster, where the representing element is the most similar one to the processed log line and not to the first cluster it matches. Moreover, the length of the shorter log line $sl$ must have at least $x\%$ length of the longer compared log line $ll$ (cf. Eq. (\ref{eqbed1})), where $x$ is the chosen similarity threshold, which specifies how similar two lines have to be to match the same cluster. Also the length of the calculated alignment must have at least the length of the shorter log line $sl$ (cf. Eq. (\ref{eqbed2})) and at least $x\%$ of the longer log line $ll$ (cf. Eq. (\ref{eqbed3})). This ensures a sequence alignment as long as possible.
\begin{align}
	\label{eqbed1}length(sl)>=x\cdot length(ll)\\
	\label{eqbed2}length(sl)+length(gaps)=length(alignment)\\
	\label{eqbed3}length(alignment)=x\cdot length(ll)
\end{align}

In the third part of the evaluation, we apply a python script for retranslating the amino acid sequences into readable UTF-8 coded text data. Therefore, as described in Sect. \ref{clustering}, the ID which is assigned to every amino acid sequence during the re-coding process is used to look up the log lines in the original log file. 

\subsection{Testdata generation}

Our test environment consisted of virtual servers running on Apache Web server and the MANTIS Bug Tracker System\footnote{https://www.mantisbt.org/} on top, a MySQL database, a firewall and a reverse proxy. The log messages of these systems are aggregated using syslog. To evaluate the presented approach, we used log data from this system. For generating the data, we applied a slightly modified version of the approach presented in \cite{skopik_semi-synthetic_2014}. With this method it is possible to generate log files of any size/time interval for a given system by simulating user input in virtual machines. In our case, we created four user machines that exhibit a typical behavior on a bug tracker system, for example, logging in and out, submitting and editing bug reports. This allowed us to control the complexity of the scenarios, inject attacks at known points of time. With this method, highly realistic conditions can be achieved. Since the deployed environment is also used in similar settings by real companies for managing bugs in their software, the produced log data is representative.

For evaluating the proposed approach, we generated 4 different log files. In order to simulate different levels of complexity, we implemented two configurations - configuration I (low complexity: the virtual users only click on the same three pages, in the same order)  and configuration II (high complexity, see \cite{skopik_semi-synthetic_2014}).
\begin{table}
	\center
	\scriptsize
	\begin{tabular}[c]{| p{0.12\columnwidth} | p{0.15\columnwidth} | p{0.13\columnwidth} | p{0.14\columnwidth} | p{0.2\columnwidth} |}
		\hline
		\centering\textit{Data Set} & \textit{Simulated Users} & \textit{Recorded Time (h)} & \textit{Data Set Length (lines)} & \textit{Used Configuration} \\ \hline\hline
		\centering U1C1 & \centering 1 & \centering 10 & \centering 484.239 & Config I\\ \hline
		\centering U4C1 & \centering 4 & \centering 10 & \centering 1.887.824 & Config I\\ \hline
		\centering U1C2 & \centering 1 & \centering 10 & \centering 413.106 & Config II\\ \hline
		\centering U4C2 & \centering 4 & \centering 10 & \centering 1.600.217 & Config II\\ \hline
	\end{tabular}
	\caption[Semi-synthetic log files]{\label{tab:semisynthtestdata}Properties of the exploited semi-synthetic log files.}
\end{table}
For generating the log files, the user activity was logged for $10$ hours. Table \ref{tab:semisynthtestdata} shows that the data set length, i.e., number of log lines, is mostly effected by the number of simulated users. In both cases (running one virtual user and running four concurrent virtual users), changing from configuration I to configuration II generated around $15\%$ less log lines. This happens because in configuration II there are more options for the virtual users to choose their actions from and there are more actions which raise a longer waiting time until a virtual user performs his next action.

\subsection{Outlier detection}\label{sect:outlierdetectioneval}
\subsubsection{Detection Capability}
Since the proposed approach has to be considered as work in progress and not all parts are fully implemented yet, we evaluated the detection capability in the context of a simple but catchy scenario. Therefore, we implemented an insider attacker who is an employee of an organization using the MANTIS bug tracker platform. The employee has valid credentials to log into the platform. Usually he accesses the database through an application hosted on the Web server, but because of a misconfiguration he found out, which port allows direct access to the database. Additionally, he uses a private device to get access to the database to steal data for unauthorized use. In our scenario, the employee wants to access one specific database entry, which he would not be authorized to access, when connecting to the database through the Web server. Therefore, when he connects to the database, a different IP address and a different MAC address, which only occurs once when accessing the database, are logged and can be detected as outlier. For simulating the scenario, we modified the log lines which are part of one logged data base access from the original log files and added it at a random location. For our proposed approach, the order of the log lines makes no difference, because they are sorted by their length, starting with the longest log line, before clustering. The log line, which includes the important information about the MAC address and the IP address is shown in List. \ref{lst:modlog}. The modified MAC and IP address are chosen randomly. 
\begin{figure}
	\begin{lstlisting}[mathescape,language=XML,columns=fullflexible,keepspaces=true,frame=tb,basicstyle=\tiny\ttfamily,numbersep=3pt,caption=Log line in which the MAC and IP address are logged during a data base access; the `*' symbols mark the parts of the log line which are modified.,captionpos=b,stringstyle=\ttfamily,label={lst:modlog},breaklines=true,keywordstyle=\bfseries,morekeywords={}]
	Jul 16 08:47:32 v3ls1316.d03.arc.local kernel: [757325.314310] iptables:ACCEPT-INFO IN=eth0 OUT= MAC=00:50:56:9c:25:67:**:**:**:**:**:**:08:00 SRC=***.***.***.*** DST=169.254.0.2 LEN=60 TOS=0x00 PREC=0x20 TTL=59 ID=36376 DF PROTO=TCP SPT=38947 DPT=80 SEQ=901703914 ACK=0 WINDOW=29200 RES=0x00 SYN URGP=0 OPT (020405B40402080A1D6066F20000000001030307) 
	\end{lstlisting}
\end{figure}

It also would be possible to detect this kind of attack with a common whitelist approach (i.e., explicitly specify the known good IP addresses and MAC addresses). However this simple, but catchy scenario allows us to show the sensitivity of our proposed approach and prove its detection capability. Furthermore, the information gathered from this elementary test scenario serves as basis for more complex cases and more complex application possibilities such as time series analysis.

To evaluate the detection capability of the model, we added the modified log lines to all the four log files mentioned in Tab. \ref{tab:semisynthtestdata}. In this evaluation, we defined clusters consisting of only one log line as outliers. To show the detection capability of the proposed model, we calculated two statistics. First, we calculated the absolute number of false positives $FP$. We defined every cluster consisting of only one log line and not including the modified version of the log line shown in List. \ref{lst:modlog} as $FP$. Second, we calculated the ratio $FPR$ between the number of $FP$ and the log file length (cf. Eq. (\ref{eq:fpr})).
\begin{align}\label{eq:fpr}
	FPR=\frac{FP}{length(\text{Log File})}
\end{align}

For the evaluation, we ran the proposed algorithm on the test data varying the similarity threshold, applied for comparing the log lines, between $85\%$ and $99\%$, raising it by $1\%$ every run. Table \ref{tab:detectthreshold} shows the similarity threshold for both methods at which the outlier we searched for was detected first. The outlier was also detected for all higher similarity thresholds. Table \ref{tab:resv1} summarizes some of the results for re-coding the log data into $L_{bio}^{full}$, which is a translation without loss of information, and Tab. \ref{tab:resv2} presents the results for re-coding the log data into $L_{bio}$, which is a translation that compresses the amount of data, but leads to a loss of information (cf. Tab. \ref{tab:rec-ex}).

Table \ref{tab:detectthreshold} indicates the lowest similarity threshold for both re-coding methods and the four test datasets at which the outlier we searched for was detected. Table \ref{tab:detectthreshold} demonstrates that a lower threshold can be chosen, when re-coding the log data to $L_{bio}$, to detect the outlier. Furthermore, the lowest threshold at which the outlier is detected is independent from the number of users and the chosen complexity of the logged network environment. It only depends on the applied re-coding model. Moreover, Tab. \ref{tab:detectthreshold} reveals that re-coding to $L_{bio}$ is more sensitive for detecting outliers since the outlier is detected at a lower threshold. This can be explained by the fact that when re-coding to $L_{bio}^{full}$ every symbol is re-coded into two symbols of the canonical amino acids alphabet. For the first symbol, only at most $13$ out of $20$ letters are used (cf. Sect. \ref{sect:recoding}). Hence, one of the first symbols occurs more often, than one of the second symbols.

In contrast to the lowest threshold at which the outlier is detected, Tab. \ref{tab:resv1} and Tab. \ref{tab:resv2} show that the number of $FP$ and also the $FPR$ depend on the complexity of the logged network environment. The $FP$ and $FPR$ is higher for the more complex configuration. According to the results, the number of logged users only has a very low influence on the number of $FP$ and the $FPR$. That was to be expected, because every user can carry out the same actions. Furthermore, the tables show that the number of $FP$ and the $FPR$ for re-coding to $L_{bio}^{full}$ (cf. Tab \ref{tab:resv1}) are a bit lower than for re-coding to $L_{bio}$ (cf. \ref{tab:resv2}). This can be explained in the same way, as in the previous paragraph due to the fact that the lowest threshold at which the outlier is detected is lower when recoding to $L_{bio}$. Again this results from the fact that re-coding to $L_{bio}$ allows a more sensitive outlier detection than re-coding to $L_{bio}^{full}$. Furthermore in both cases, the $FP$ and $FPR$ start increasing much faster at a specific threshold. Again, because of the higher sensitivity this can be recognized earlier, when re-coding to $L_{bio}$ (at $93\%$ similarity) than when re-coding to $L_{bio}^{full}$ (at $96\%$). Thus, when using a higher similarity threshold, which generates more $FP$, the outliers can be clustered again, applying a lower threshold. This makes it easier to understand the detected outliers and increase situational awareness, since they are grouped. Using realistic similarity thresholds, especially the thresholds summarized in Tab. \ref{tab:detectthreshold}, the number of $FP$ and the $FPR$ are very low and the detected outliers can be easily investigated manually by a system administrator.

\begin{table*}[tb]
	\centering
	\scriptsize
	\caption{$FP$ and $FPR$ results, when recoding to  $L_{bio}^{full}$}
	\label{tab:resv1}
	\begin{tabular}{c|rrrrrrrr}
		\hline
		Threshold & \multicolumn{1}{c}{$FP_{U1C1}$} & \multicolumn{1}{c}{$FPR_{U1C1}$} & \multicolumn{1}{c}{$FP_{U1C2}$} & \multicolumn{1}{c}{$FPR_{U1C2}$} & \multicolumn{1}{c}{$FP_{U4C1}$} & \multicolumn{1}{c}{$FPR_{U4C1}$} & \multicolumn{1}{c}{$FP_{U4C2}$} & \multicolumn{1}{c}{$FPR_{U4C2}$} \\ \hline
0.86 & 14 & 2,89E-05 & 202 & 4,89E-04 & 4 & 2,12E-06 & 160 & 1,00E-04 \\
0.88 & 17 & 3,51E-05 & 286 & 6,92E-04 & 6 & 3,18E-06 & 378 & 2,36E-04 \\
0.91 & 24 & 4,96E-05 & 696 & 1,68E-03 & 13 & 6,89E-06 & 1718 & 1,07E-03 \\
0.92 & 27 & 5,58E-05 & 758 & 1,83E-03 & 16 & 8,48E-06 & 2052 & 1,28E-03 \\
0.95 & 45 & 9,29E-05 & 1659 & 4,02E-03 & 40 & 2,12E-05 & 4955 & 3,10E-03 \\
0.96 & 2223 & 4,59E-03 & 5397 & 1,31E-02 & 2973 & 1,57E-03 & 11978 & 7,49E-03 \\
0.97 & 16972 & 3,50E-02 & 25763 & 6,24E-02 & 59289 & 3,14E-02 & 87107 & 5,44E-02 \\ \hline
	\end{tabular}
\end{table*}

\begin{table*}[tb]
	\centering
	\scriptsize
	\caption{$FP$ and $FPR$ results, when recoding to  $L_{bio}$}
	\label{tab:resv2}
	\begin{tabular}{c|rrrrrrrr}
		\hline
		Threshold & \multicolumn{1}{c}{$FP_{U1C1}$} & \multicolumn{1}{c}{$FPR_{U1C1}$} & \multicolumn{1}{c}{$FP_{U1C2}$} & \multicolumn{1}{c}{$FPR_{U1C2}$} & \multicolumn{1}{c}{$FP_{U4C1}$} & \multicolumn{1}{c}{$FPR_{U4C1}$} & \multicolumn{1}{c}{$FP_{U4C2}$} & \multicolumn{1}{c}{$FPR_{U4C2}$} \\ \hline
		0.86 & 15 & 3,10E-05 & 362 & 8,76E-04 & 15 & 7,95E-06 & 483 & 3,02E-04 \\
		0.87 & 18 & 3,72E-05 & 523 & 1,27E-03 & 16 & 8,48E-06 & 901 & 5,63E-04 \\
		0.88 & 22 & 4,54E-05 & 701 & 1,70E-03 & 19 & 1,01E-05 & 1878 & 1,17E-03 \\
		0.90 & 33 & 6,81E-05 & 825 & 2,00E-03 & 24 & 1,27E-05 & 2119 & 1,32E-03 \\
		0.92 & 48 & 9,91E-05 & 1160 & 2,81E-03 & 41 & 2,17E-05 & 3363 & 2,10E-03 \\
		0.93 & 523 & 1,08E-03 & 2388 & 5,78E-03 & 521 & 2,76E-04 & 6030 & 3,77E-03 \\
		0.94 & 8852 & 1,83E-02 & 13964 & 3,38E-02 & 21308 & 1,13E-02 & 35144 & 2,20E-02 \\ \hline
	\end{tabular}
\end{table*}

\begin{table}[tb]
	\centering
	\scriptsize
	\caption{Thresholds at which the outlier is detected.}
	\label{tab:detectthreshold}
	\begin{tabular}{c|rr}
		\hline
		Conf. & \multicolumn{1}{c}{$L_{bio}^{full}$} & \multicolumn{1}{c}{$L_{bio}$} \\ \hline
		U1C1 & 0,91 & 0,88 \\
		U1C2 & 0,91 & 0,87 \\
		U4C1 & 0,92 & 0,86 \\
		U4C2 & 0,92 & 0,88\\ \hline
	\end{tabular}
\end{table}

\subsubsection{Model Scalability}

\begin{figure}[!tb]
  \centering
  \subfigure[total time.]{
    \label{fig:a}
    \resizebox{0.46\linewidth}{!}{\includegraphics{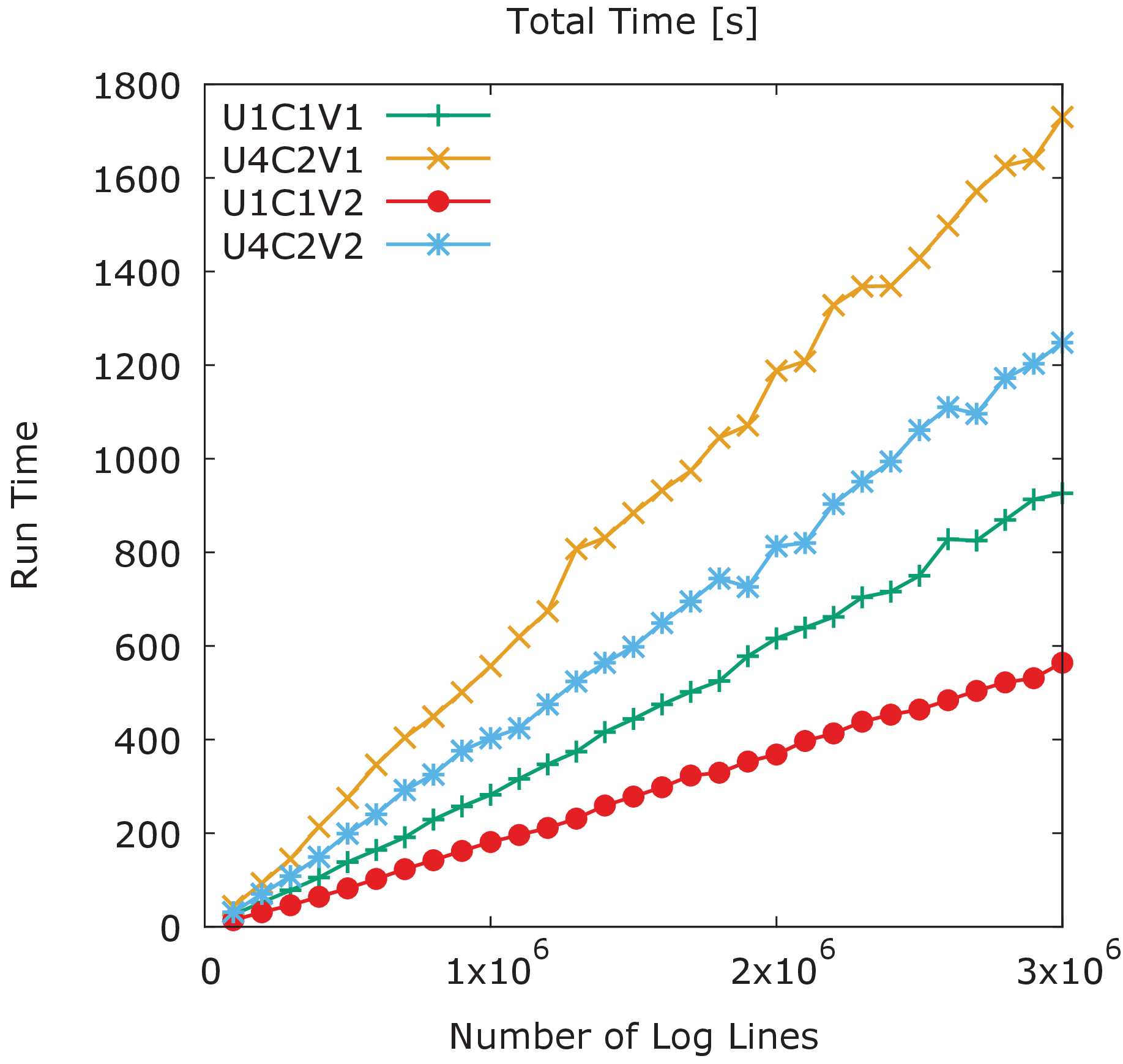}}
  }
  \subfigure[re-coding time.]{
    \label{fig:b}
    \resizebox{0.46\linewidth}{!}{\includegraphics{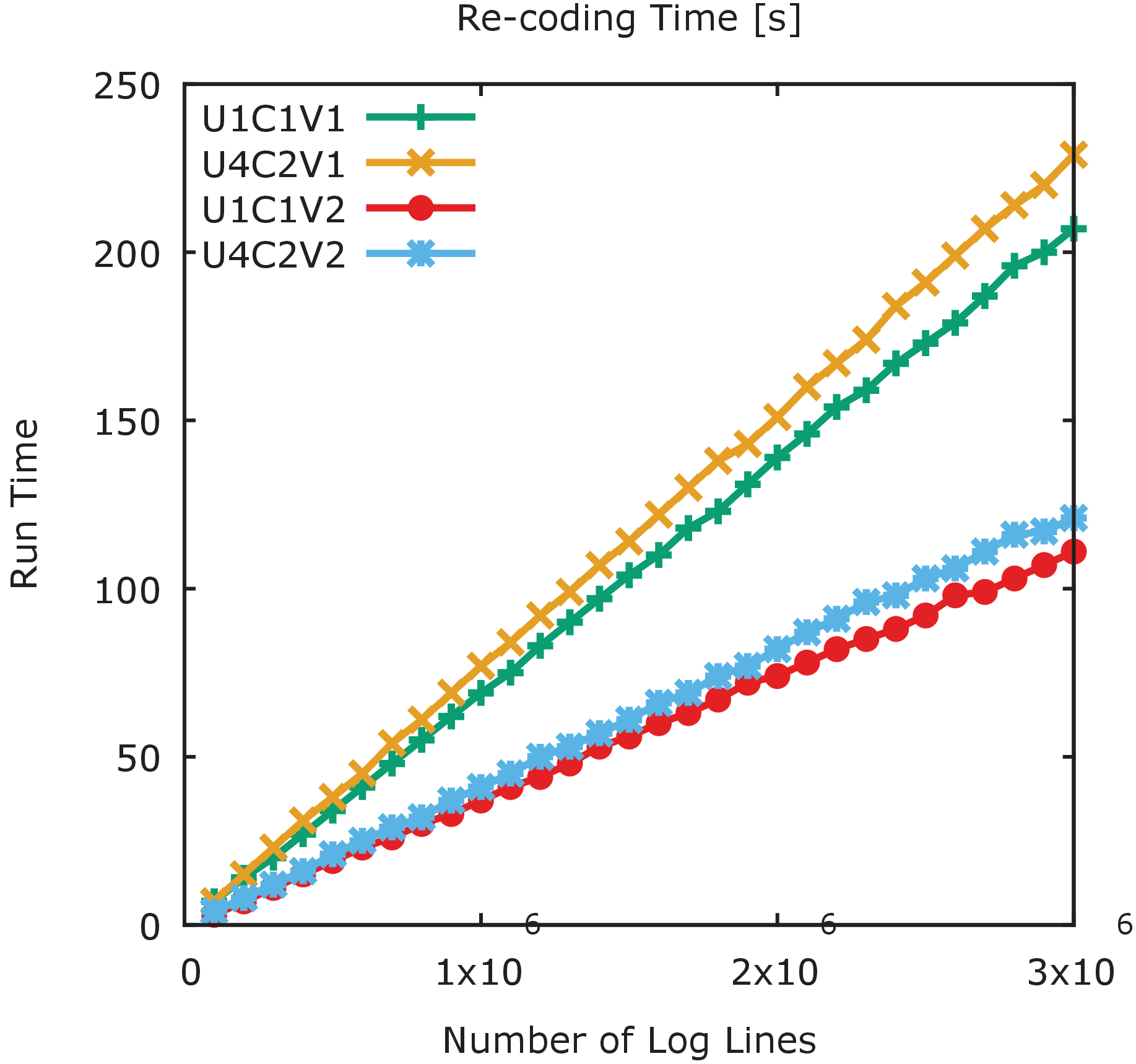}}
  }  
  \subfigure[clustering time]{
    \label{fig:c}
    \resizebox{0.46\linewidth}{!}{\includegraphics{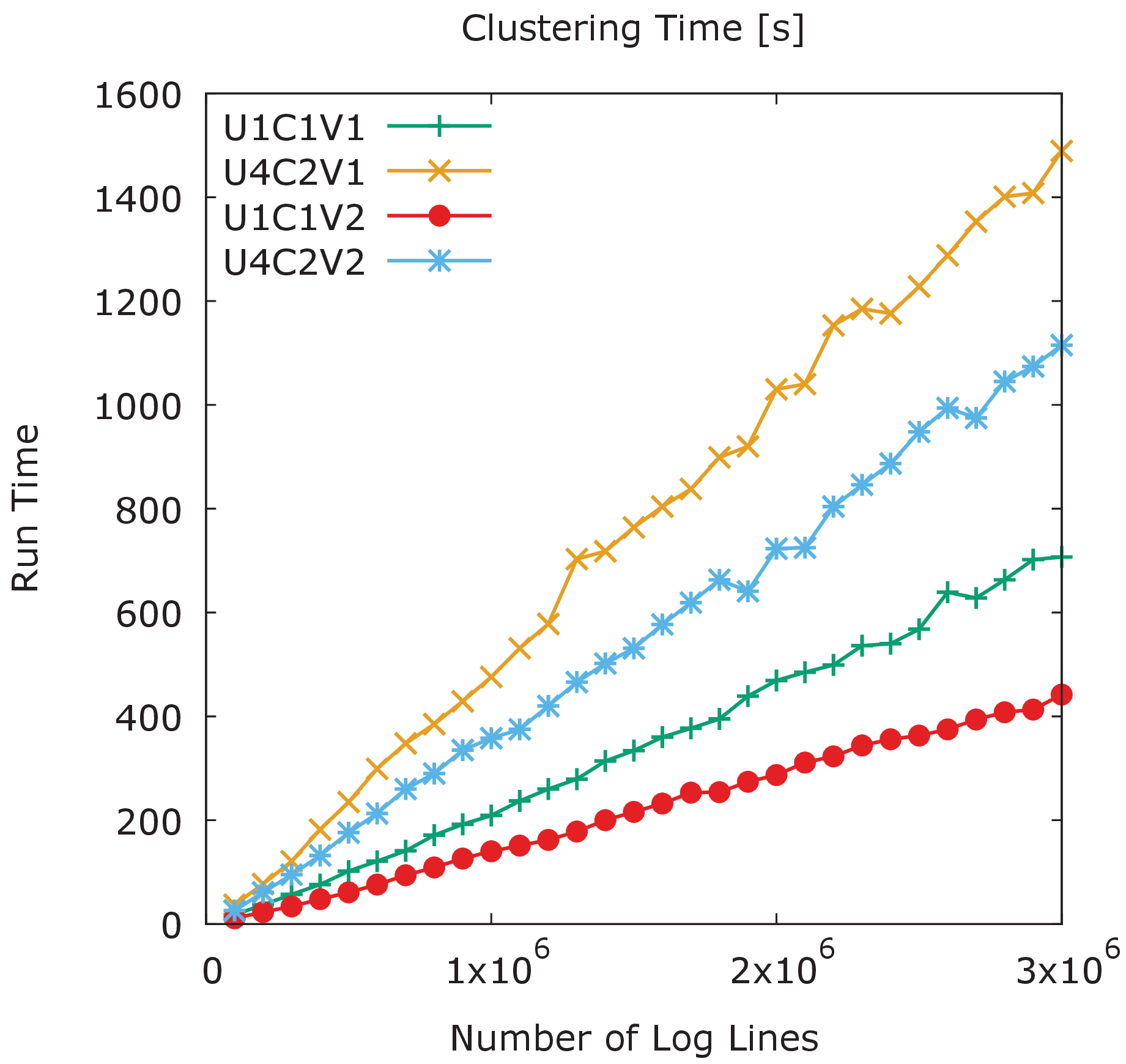}}
  }
  \subfigure[re-translation time.]{
    \label{fig:d}
    \resizebox{0.46\linewidth}{!}{\includegraphics{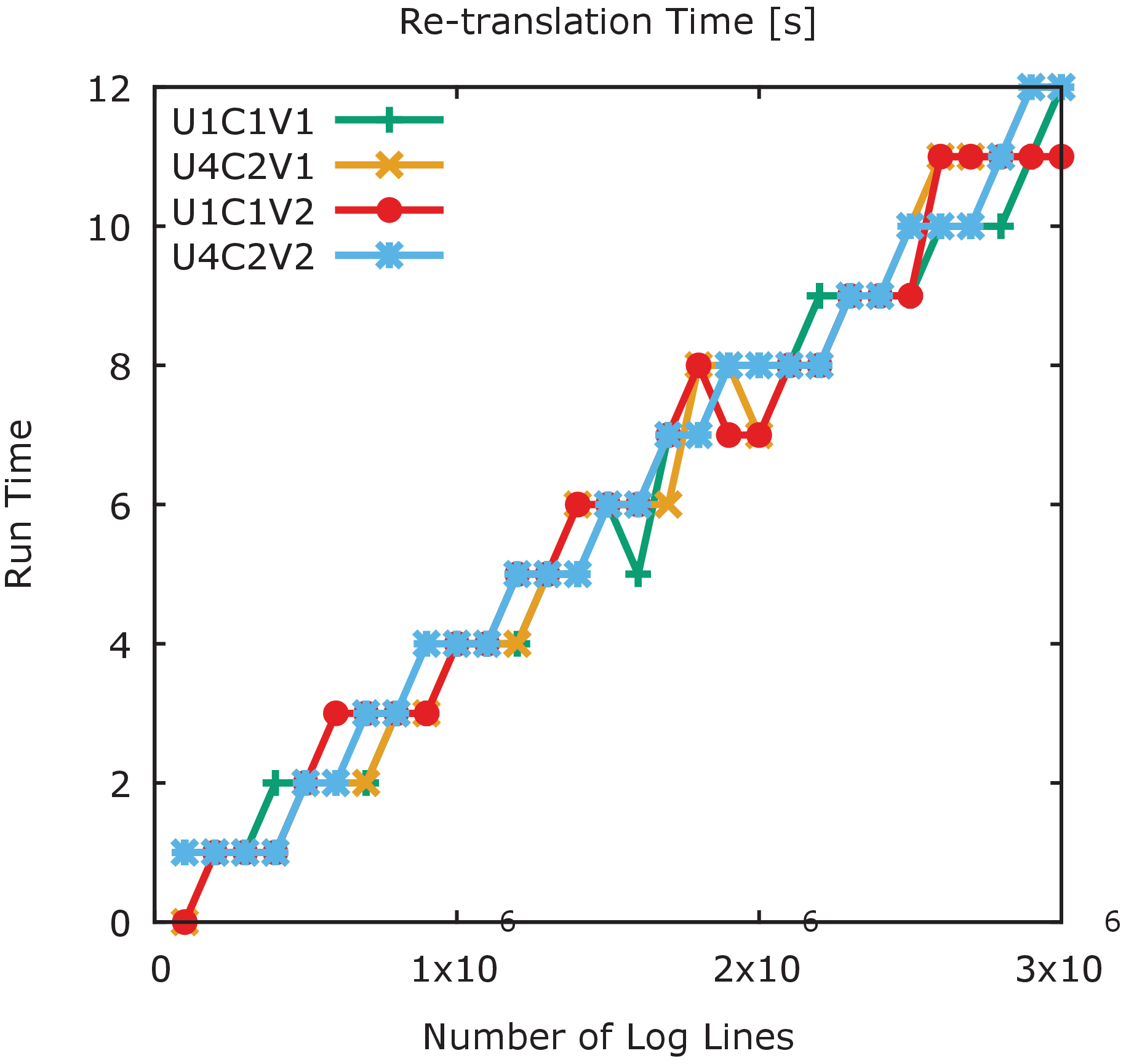}}
  }  
  \caption{\label{fig:runtime}Run time and scalability for the single steps of the proposed approach.} 
\end{figure}

To evaluate the scalability of our approach for detecting outliers, we generated log files of different lengths, i.e., line numbers, for the simplest configuration $U1C1$ and the most complex configuration $U4C2$, cf. Tab. \ref{tab:semisynthtestdata}. The length of the log files ranges from $100,000$ lines to $3,000,000$ lines. For generating the log files, we applied the approach proposed in \cite{wurzenberger_complex_2016}. This algorithm allows to generate highly realistic semi-synthetic log files based on a small piece of real log data. We compared the run-time for re-coding to $L_{bio}^{full}$ (V1) and re-coding to $L_{bio}$ (V2). As similarity threshold, we used the values obtained from the analysis of the lowest value at which the outlier was detected (cf. Tab. \ref{tab:detectthreshold}). Besides the total run time we calculated the time for re-coding, clustering and re-translating.  The plots in Fig. \ref{fig:runtime} demonstrate that the runtime of the model is increasing linearly. The total runtime is higher for the more complex configuration and also for the translation to $L_{bio}^{full}$. Depending on the complexity of the data, in our test environment (c.f. Sect. \ref{evalconf}) the algorithm is able to process between $1800$ and $5000$ log lines per second. The re-coding time only depends on the re-coding method and is longer for recoding to $L_{bio}^{full}$, since more symbols are generated. The clustering time depends on the complexity of the analyzed system and on the re-coding method, i.e., on the length of the analyzed sequences. The re-translation time only depends on the length of the log file. The most time consuming part is the clustering.
  
\subsection{Outlook on further application possibilities}\label{outlook}

The outlined approach is in an early stage and not all parts of the proposed model are fully implemented yet. There exist also other application possibilities of the proposed approach than evaluated in Sect. \ref{sect:outlierdetectioneval}. To enable real-time log data processing and outlier detection, we foresee the application of the following method: First the clustering model is trained with log data of optional length, which represents the normal system behavior of the monitored network environment. The training log data should at least cover a cycle, which includes also activities such as update and back up processes, which are usually done in specific time periods. Afterwards new log lines obtained from the system are sorted to the clusters. If a log line does not match to any cluster it is considered as outlier and raises an alarm. Furthermore, if the system administrator decides that a detected outlier does not represent anomalous behavior, the log line can be added to the cluster model as representative element of a new cluster. Since the training phase to generate new clusters just runs occasionally (and potentially in parallel to the regular detection of outliers), its run-time does not negatively influence the actual detection. In this phase, also clusters with just one member do not represent outliers, but are filled up in the later detection phase with log line instances. This means, in the training phase a higher similarity threshold can be set. This would especially decrease the number of $FP$ and raise the detection capability.

The proposed model can also be applied for time series analysis to detect attacks and invaders. For this purpose, clustering models are created for different time periods. Then, the properties of the obtained clustering models can be compared (e.g., between two consecutive hours or days). If one compares two clustering models and one cluster occurs only in one of the two models, these cluster can be seen as outliers (e.g., a new device was plugged to the network, which should not be there). Furthermore also a change in the size of a cluster is an indication for anomalous behavior. For example if an attacker ex-filtrates data, the number of log lines referring to the database server will increase (especially in relation to the number of log lines in other clusters). If this is done with a machine, which belongs to the company and therefore uses a legitimate MAC and IP address the log lines would not be recognized as outliers. But in the time series analysis it would be clearly visible that the sizes of specific clusters, which are related to the database server, are increasing, and thus a major change in the system utilization behavior detected.
\section{Conclusion and Future Work}
\label{sec:conclusion}

This paper describes a novel model, which allows to apply high-performance bioinformatics tools in the context of anomaly detection on log data produced in computer networks. Since most of the bioinformatics tools operate on canonical amino acid sequences, we introduced two different methods to re-code log data coded in UTF-8 code, consisting of $255$ symbols, to the alphabet of canonical amino acids, consisting of only $20$ symbols. The first method describes a translation without loss of information, while the second method describes a translation, which compresses the data and therefore some information gets lost, but it allows faster anomaly detection. We further demonstrated how the re-coded log data can be clustered applying bioinformatics algorithms and tools for generating sequence alignments. We furthermore explained how the output can be re-translated into a human-readable format using an ID number. Finally, we described and evaluated the outlier detection.

In opposite to most other approaches, which work with word matching algorithms, our model implements a character-based sequence comparison. This allows a much more sensitive anomaly detection (e.g., similar URLs with slight deviations are recognized as related). Since the bioinformatics tools are developed for many years they are optimized for high-performance and high data throughput to allow processing huge amounts of data in very short times. Furthermore, our approach does not have to know about syntax and semantics of the log data (i.e., no specific parsers are required to detect anomalous system behavior). Hence, it can be applied in any computer network, which logs events in text formats. This especially enables the application in legacy systems (e.g., in the Industrial Control Systems (ICS) domain, which are often not well documented, as well as in less mature systems with a small market share).

In the future, we plan to focus on further development of the re-coding model described in Sect. \ref{sect:recoding} and the scoring system defined in Sect. \ref{sect:compare2lines}. Therefore, we intend to modify the re-coding function, so that often re-occurring parts of log files (e.g., static texts) are translated into less symbols and therefore accept a higher loss of information for these parts; and less frequent parts, which include the more interesting variable parts (e.g., IP addresses, user names, port numbers) of log lines should be translated without loss of information. Furthermore, we plan to investigate if the scoring system can be adjusted, so that, similar to the analysis of amino acid sequences, the score is higher for highly related letters and lower in other cases. Moreover, we want to implement and evaluate the application methods of our approach described in Sect. \ref{outlook} for real-time outlier detection and time series analysis.


\ifCLASSOPTIONcompsoc
  \section*{Acknowledgments}
This work was partly funded by the FFG project synERGY (855457) and carried out in course of a PhD thesis at the Vienna University of Technology funded by the FFG project BAESE (852301).



\bibliographystyle{IEEEtran}
\bibliography{refs}
%
%
%

\end{document}